\documentclass{article}
\usepackage{spconf,amsmath,graphicx}
\usepackage{appendix}

\title{Spatial Equalization Before Reception: Reconfigurable Intelligent Surfaces for Multi-path Mitigation \\ \normalsize{(Invited Paper)}}
\name{Hongliang Zhang$^{\dagger}$, Lingyang Song$^{\star}$, Zhu Han$^{\star\dagger}$, and H. Vincent Poor$^{\dagger}$}
\address{$^{\dagger}$Department of Electrical Engineering, Princeton University, Princeton, NJ, USA.\\
$^{\star}$Department of Electronics, Peking University, Beijing, China.\\	
$^{\star\dagger}$Electrical and Computer Engineering Department, University of Houston, Houston, TX, USA.
\\}
\begin{document}
%
\newtheorem{remark}{\bf Remark}
\maketitle
\begin{abstract}
Reconfigurable intelligent surfaces (RISs), which enable tunable anomalous reflection, have appeared as a promising method to enhance wireless systems. In this paper, we propose to use an RIS as a spatial equalizer to address the well-known multi-path fading phenomenon. By introducing some controllable paths artificially against the multi-path fading through the RIS, we can perform equalization during the transmission process instead of at the receiver, and thus all the users can share the same equalizer. Unlike the beamforming application of the RIS, which aims to maximize the received energy at receivers, the objective of the equalization application is to reduce the inter-symbol interference (ISI), which makes phase shifts at the RIS different. To this end, we formulate the phase shift optimization problem and propose an iterative algorithm to solve it. Simulation results show that the multi-path fading effect can be eliminated effectively compared to benchmark schemes.   
\end{abstract}
\vspace{-2mm}
\section{Introduction}
\label{sec:intro}

An increase in the number of mobile devices in the past decade has highlighted the need for high-speed data services in future wireless communication systems. Although various technologies have been developed to strengthen target signals such as relays and multiple-input multiple-output (MIMO) systems, network operators have been continuously struggling to build wireless networks that can guarantee to provide high quality-of-service (QoS) in the presence of harsh wireless propagation environments due to uncontrollable interactions of transmitted waves with surrounding objects and their destructive interference at receivers~\cite{basar2019reconfigurable}. 

Fortunately, recent developments of meta-material have given a rise to a new opportunity to enable the control of wireless propagation environments \cite{di2020smart}. In particular, the use of reconfigurable intelligent surfaces~(RISs), consisting of ultra-thin meta-materials inlaid with multiple sub-wavelength scatters, has emerged as a cost-effective solution to create favorable propagation environments \cite{zhang2020}. This can be achieved by controlling phase shifts of impinging radio waves at the RIS such that incident signals can be reflected towards intended receivers \cite{el2020reconfigurable}. 

In the literature, RIS-aided wireless communications have attracted considerable interest. Particularly, previous studies focused on the application of RISs for beamforming, which aims to maximize the data rate by adjusting phase shifts at the RIS. In \cite{yu2019miso}, the data rate of a point-to-point RIS-assisted multi-input single-output (MISO) system was maximized by jointly optimizing the beamformer at the transmitter and continuous phase shifts of the RIS. The authors in \cite{zhang2020reconfigurable} derived the achievable data rate and discussed how a limited number of phase shifts influences the data rate in a point-to-point RIS-assisted communication system. The authors in \cite{di2020hybrid} proposed a hybrid beamforming scheme for a multi-user RIS-assisted MISO system together with a phase shift optimization algorithm to maximize the sum-rate. In \cite{gao2020reconfigurable}, the data rate in RIS-aided multi-user MISO systems was maximized by jointly considering power allocation and phase shift at the RIS with user proportional fairness constraints.

In this paper, unlike the above noted works, we propose to exploit the potential of the RIS as a spatial equalizer to address multi-path fading. To be specific, we consider a downlink multi-user MISO communication system, where some controllable paths are introduced via the RIS to combat multi-path fading. Different from traditional communication systems, where equalization can only be done at receivers, the proposed scheme can achieve equalization in the transmission process, and thus multiple users can share the same RIS which is more cost-effective. However, since the objective of the spatial equalizer is to reduce the inter-symbol interference (ISI), the phase shift design of the RIS for beamforming applications cannot be applied directly. To this end, we formulate the ISI minimization problem by optimizing the phase shifts at the RIS and propose an algorithm to solve this problem efficiently. Simulation results verify the effectiveness of the RIS based spatial equalizer, and how the size of the RIS impacts the performance is also discussed.  

 \begin{figure}[!t]
	\centering
	\includegraphics[width=3.0in]{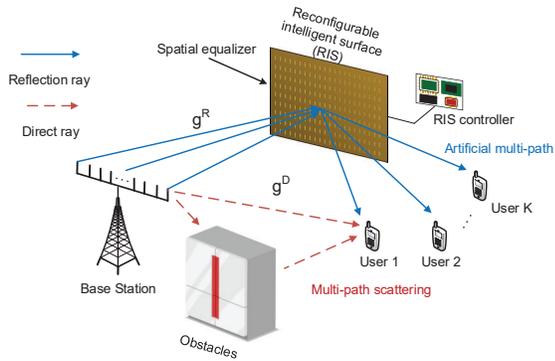}
	\vspace{-3mm}
	\caption{System model for the RIS-assisted spatial equalization.}
	\vspace{-4mm}
	\label{scenario}
\end{figure}

\vspace{-2mm}
\section{System Model}
\label{sec:system}

As shown in Fig. \ref{scenario}, we consider a downlink multi-user RIS-assisted MISO communication network consisting of one base station (BS) with $M$ antennas and $K$ single-antenna users, denoted by $\mathcal{K} = \{1,\ldots, K\}$. To reduce the ISI, an RIS is deployed as a spatial equalizer. The RIS is composed of $N$ electrically controllable elements with the size length being $a$, denoted by $\mathcal{N} = {1,\ldots,N}$.  Each element can adjust its phase shift by switching Positive-Intrinsic-Negative (PIN) diodes between ``ON" and ``OFF" states. Due to some physical limitations, the state transition for each PIN diode may cost some time. In this paper, within a considered period, we assume that the phase shift for each element is fixed. Define $\theta_n$ as the phase shift for element $n$, and the reflection factor of element $n$ can be written by $\Gamma_n = \Gamma e^{-j\theta_n}$, where $\Gamma \in [0,1]$ is a constant.

For each user, it can receive two rays of signals. The first ray is the direct link from the BS, which consists of the scattered signals from the environment. We define $g^{D}_k(t)$ as the channel impulse response of the direct link from the BS to user $k$, which models independent fast fading and path loss. To be specific, $g^{D}_k(t)$ can be written as
\begin{equation}
	g^{D}_k(t) = (\beta_k^D)^{1/2}h^{D}_k(t),
\end{equation}
where $h^{D}_k(t)$ is the fast fading coefficient caused by the multi-path effect and $\beta_k^D$ is the path loss related to distance $d_k$ between the BS and user $k$, i.e., $\beta_k^D = Gd_k^{-\alpha}$. Here, $G$ is a normalized factor for the direct link and $\alpha$ is the path loss exponent. 

The second ray is the reflection link through the RIS. Each RIS element will reflect the incident signals from the BS to users to eliminate the multi-path effect. We define $g_{n,k}^R(t)$ as the channel impulse response of the reflection link through RIS element $n$ to user $k$, which also includes independent fast fading and path loss. Specifically, $g^{R}_{n,k}(t)$ can be written as
\begin{equation}
	\vspace{-1mm}
	g^{D}_{n,k}(t) = (\beta_{n,k}^R)^{1/2}\Gamma_nh^{R}_{n,k}(t),
\end{equation}
where $h^{R}_{n,k}(t)$ is the fast fading coefficient and $\beta_{n,k}^R$ is the path loss related to distance $l_{n}$ between the BS and the $n$-th RIS element, and distance $l_{n,k}$ between the $n$-th RIS element and user $k$. According to the result in \cite{zeng2020reconfigurable}, we have $\beta_{n,k}^R = G'(l_{n}l_{n,k})^{-\alpha}$ where $G'$ is a normalized factor for the reflection link. It is worthwhile to point out that we can approximate the distance to different RIS elements as the distance to the center of the RIS when $l_n, l_{n,k} \gg a$ \cite{zhang2020reconfigurable}. Therefore, we have $\beta_{n,k}^R \approx \tilde{\beta}_{k}^R, \forall n \in \mathcal{N}$, where $\tilde{\beta}_{k}^R$ is the path loss of the link going through the center of the RIS.

Define one-bit signal for user $k$ as $s_k(t)$, and the received signal at user $k$ can be written as\footnote{In this paper, we assume that perfect beamforming is done by the BS, and thus the interference between different users are neglected here. The beamformer design is not studied due to the length limit.}
\begin{equation}
		\vspace{-1mm}
	y_k(t) = \left(g_k^D(t) + \sum\limits_{n \in \mathcal{N}}g_{n,k}^D(t)\right)*s_k(t),
		\vspace{-1mm}
\end{equation}
where $*$ is the convolution operator.
\vspace{-2mm}
\section{Problem Formulation}
The objective of this paper is to reduce ISI through the RIS-based spatial equalizer. In the following, we will first introduce how to extract ISI using the peak distortion analysis and formulate the ISI minimization problem.

\textbf{ISI Extraction:} Assuming that $y_k(t)$ achieve its maximum at $t = 0$ and $T$ is the sampling interval for one bit. According to \cite{casper2002accurate}, the ISI for user $k$ can be written as 
\begin{equation}
	\vspace{-1mm}
I_k = \sum\limits_{i = -\infty,i \ne 0}^{\infty} y_k (t - iT)|_{t=0},
	\vspace{-1mm}
\end{equation}
under the assumption that only one bit is transmitted. In practice, we will only considered the ISI within a window.

\textbf{Problem Formulation:} Note that the RIS is not equipped with any delay components and thus cannot control the spread of multi-paths. In practice, we will select a boundary which includes most significant ISI for the ease of ISI calculation. Therefore, the objective of the spatial equalizer is to reduce the energy of remaining ISI within the considered boundary after equalization. In consideration of the fairness, we will minimize the maximum power of ISI among these users by adjusting phase shifts at the RIS. Mathematically, the optimization problem can be written as
\begin{subequations}
	\begin{align}
 (P1): ~~~	\min \limits_{\{\theta_n\}, \eta} ~~& \eta, \\
	   s.t. ~~& I_k I^{*}_k \leq \eta, \forall k \in \mathcal{K},\label{postive} \\
	   & \eta \geq 0,
	\end{align}
\end{subequations}
where $\eta$ is the maximum power of ISI among these users, and $I^{*}_k$ is the conjugate of $I_k$.
\vspace{-2mm}
\section{Algorithm Design}
\label{sec:algorithm}

In this section, we will propose a phase shift optimization (PSO) algorithm to solve problem (P1) efficiently. Define $\mathcal{F}(\cdot)$ as the Fourier transformation operator. Let $H_k^D(\omega) = \mathcal{F}(g_k^D(t))$, $H_{n,k}^R(\omega) = \mathcal{F}(g_{n,k}^R(t))$, $S_{k}(\omega) = \mathcal{F}(s_{k}(t))$, and $Y_k(\omega) = \mathcal{F}(y_k(t))$. With these notations, we have 
\begin{equation}\label{freq}
		\vspace{-1mm}
	Y_k(\omega) = \left(H_{k}^D(\omega) + \sum\limits_{n \in \mathcal{N}} H_{n,k}^R(\omega)\right) S_k(\omega).
		\vspace{-1mm}
\end{equation}

According to the definition of the Fourier transformation, we have
\begin{equation}
		\vspace{-1mm}
	Y_k(0) = \int_{-\infty}^{\infty}y_k(t)dt \approx (y_k(0) + I_k)T.
		\vspace{-1mm}
\end{equation}
Therefore, we can have the following equation~\cite{song2018a}:
\begin{equation}
		\vspace{-1mm}
	I_k = \frac{Y_k(0)}{T} - y_k(0).
		\vspace{-1mm}
\end{equation}

Note that phase shifts of the RIS will not affect $y_k(0)$ as the transmission delay through the RIS is typically longer than the direct one. Motivated by this observation, we optimize $Y_k(0)$ by tuning phase shifts of the RIS. In the following, we will elaborate on how to find the optimal phase shifts.

Given $y_k(0)$, optimization problem (P1) can be solved by the Lagrange-Dual technique. Let $\mu_k$ be the Lagrange multiplier corresponding to the ISI constraint for user $k$, the Lagrangian can be written as
\begin{equation}\label{definition}
		\vspace{-1mm}
	L(\theta_n, \eta, \mu_k) = \eta + \sum\limits_{k \in \mathcal{K}}\mu_k\left( \left|\frac{Y_k(0)}{T} - y_k(0)\right|^2 - \eta\right),
		\vspace{-1mm}
\end{equation}
and the dual problem can be written as
\begin{equation}
		\vspace{-1mm}
	\max\limits_{\mu_k,\nu_k \geq 0} \min \limits_{\theta_n,\eta} L(\theta_n, \eta, \mu_k).
		\vspace{-1mm}
\end{equation}
The problem can be solved by gradient based method~\cite{boyd2004convex}. In the $l$-th iteration, primal and dual problems are solved in the following way:

\textbf{Primal Problem:} In the primal problem, we solve $\theta_n$ and $\eta$ given the value of $\mu_k$. To be specific, we have
	\begin{align}
			\vspace{-1mm}
 		\eta^{l + 1} &= [\eta^{l} - \delta_{\eta} \nabla_{\eta}^l L(\theta^l_n,\eta^l,\mu^l_k)]^{+}, \\
 		\theta_n^{l + 1} &= \theta_n^{l} - \delta_{\theta} \nabla_{\theta_n}^l L(\theta^l_n,\eta^l,\mu^l_k),
 			\vspace{-1mm}
	\end{align}
where $[a]^{+} = \max\{0,a\}$, $\delta_{\eta}$ and $\delta_{\theta}$ are step sizes of $\eta$ and $\theta_n$, respectively. Here, the gradients can be calculated by
	\begin{align}
			\vspace{-1mm}
		&\nabla_{\eta}^l L(\theta^l_n,\eta^l,\mu^l_k) = 1 - \sum\limits_{k \in \mathcal{K}}\mu_k,\\
		&\nabla_{\theta_n}^l L(\theta^l_n,\eta^l,\mu^l_k) = \nonumber\\
		&~~~~~~~~~~~~~2\sum\limits_{k \in \mathcal{K}}\frac{\mu_k}{T^2}\left(A_{k,n}B_{k,n}^{*}je^{j\theta_n^l} \hspace{-1mm}-\hspace{-1mm} B_{k,n}A_{k,n}^{*}je^{-j\theta_n^l} \right. \nonumber\\
		&~~~~~~~~~~~~~\left. y_k^{*}(0)TB_{k,n}je^{-j\theta_n^l}-y_k(0)TB^{*}_{k,n}j e^{j\theta_n^l} \right),\label{gradient}
			\vspace{-1mm}
	\end{align}
where $Y_k(0) = A_{k,n} + B_{k,n} e^{-j\theta_n}$. The detailed proof of (\ref{gradient}) is given in the Appendix.

\textbf{Dual Problem:} In the dual problem, we fix the results $\theta_n$ and $\eta$, and solve the dual variable $\mu_k$. According to \cite{boyd2004convex}, $\mu_k$ can be updated in the following way:
\begin{equation}
	\mu_k^{l + 1} \hspace{-1mm} =\hspace{-1mm} \left[\mu_k^{l} + \delta_{\mu}\left(\left|\frac{Y^{l+1}_k(0)}{T} - y_k(0)\right|^2 \hspace{-1mm}- \eta^{l+1}\right)\right]^{+}, k \in \mathcal{K},
\end{equation}
where $Y^{l+1}_k(0)$ is obtained by $\theta_n^{l+1}$ and $\delta_{\mu}$ is a step size of $\mu_k$.

 \begin{figure}[!t]
	\centering
	\includegraphics[width=2.6in]{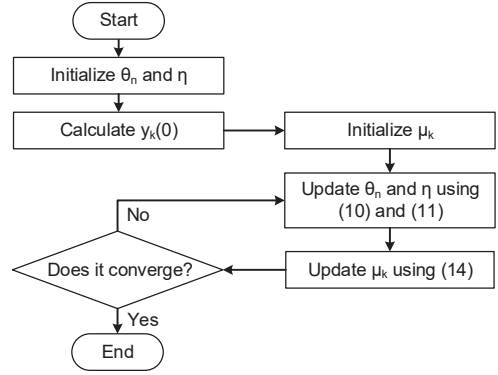}
	\vspace{-3mm}
	\caption{Flowchart of the proposed PSO algorithm.}
	\vspace{-4mm}
	\label{flowchart}
\end{figure}

The PSO algorithm can be summarized as the flowchart given in Fig.~\ref{flowchart}. In each iteration, we use the primal-dual gradient method to obtain phase shifts $\theta_n$ and the maximum power of ISI $\eta$ for all users. The termination condition is that the difference of the values of the objective for two successive iterations is less than a predefined threshold $\sigma$. It is worthwhile to point out that the obtained solution is local-optimal since the original problem is non-convex. The complexity of the proposed PSO algorithm should be $\mathcal{O}(\sqrt{K}\log(1/\sigma))$ \cite{boyd2004convex}. This implies that we can adjust the complexity according to the requirements of applications by tuning $\sigma$.
\begin{remark}\label{rem}
	When we neglect the fading and the number of RIS elements is even, the RIS filter can achieve at least the same performance with the one without the RIS in terms of the maximum ISI power. We can achieve this by setting phase shifts of two adjacent elements as 0 and $\pi$, respectively. 
\end{remark}

\vspace{-2mm}
\section{Simulation Results}
\label{sec:simulation}

In this section, we evaluate the performance of the proposed PSO algorithm. The parameters are selected according to the 3GPP standard \cite{3gpp2018study} and existing work \cite{zhang2020reconfigurable}. The height of the BS is 25m. We set the number of users $K = 4$ and the number of antennas $M = 10$. The users are uniformly located in a square area whose side length is set as 100m, and distance from the center of this area to the BS is 100m. The RIS is placed in parallel to the direction from the BS to the center of the user area, and the horizontal distance from the BS to the RIS is 100m. We assume that the distance between the center of the RIS and the projected point of the BS at the RIS-plane is $D = 50$m. The center of the RIS is located at the middle between the BS and the square area with the height being 25m. The carrier frequency is set as 5.9 GHz, the size length of an RIS element is set as $a = 0.02$m, and the number of RIS elements is set as $N = 100$. We also assume that the RIS is fully reflected, i.e., $\Gamma = 1$.  For the channel model, the path loss exponent is set as $\alpha = 2$. The normalized factor $G =  G' = -43$dB. The stochastic model in \cite{saleh1987statistical} is used to capture the multi-path effect. For the direct ray, we assume that there exist $L$ paths and each RIS element corresponds to a reflection path. The sampling interval is set $T = 1$ms. We set convergence threshold $\delta = 0.01$. All numeral results are obtained by 300 Monte Carlo simulations.

In comparison, we also present the performance of the following schemes: 1) Random phase shift (RPS) scheme: the phase shift for each RIS element is selected randomly; 2) Discrete phase shift (DPS) scheme: the phase shift for each RIS element is discrete, i.e., 2-bit quantified in this simulation. We will select the phase shift value which is closest to the solution obtained by the proposed PSO algorithm.  3) Non-RIS scheme: the spatial equalizer is removed.

 \begin{figure}[!t]
	\centering
	\includegraphics[width=2.7in]{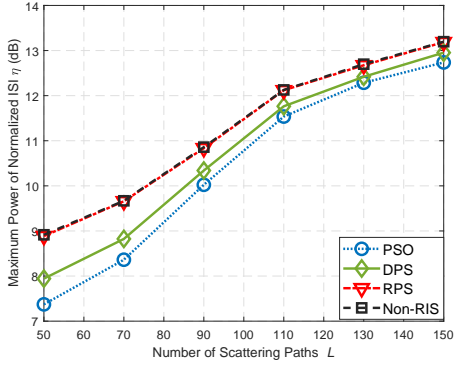}
	\vspace{-3mm}
	\caption{Maximum power of normalized ISI $\eta$ for different number of scattering paths $L$.}
	\vspace{-4mm}
	\label{comparison}
\end{figure}

In Fig.~\ref{comparison}, we present the maximum power of normalized ISI $\eta$ for different number of scattering paths $L$. Here, we normalize the received power for each user at $t = 0$ as 1. From this figure, we can observe that the proposed PSO algorithm can outperform other benchmark algorithms. We can also learn that even with 2-bit quantization at the RIS, we can reduce 1 dB compared to that without the RIS filter in terms of the maximum power of normalized ISI when $L = 100$. These observations are consistent with Remark \ref{rem}. Moreover, we can observe that random phase shifts at the RIS can achieve almost the same performance as that without the RIS. On the other hand, $\eta$ will increase as the number of scattering paths $L$ grows, and the benefit brought by the RIS filter will drop due to the limited size of the RIS. 

 \begin{figure}[!t]
	\centering
	\includegraphics[width=2.7in]{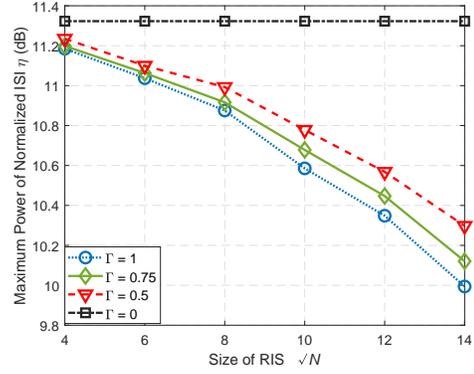}
	\vspace{-3mm}
	\caption{Maximum power of normalized ISI $\eta$ for different size of the RIS $\sqrt{N}$.}
	\vspace{-4mm}
	\label{size}
\end{figure}

In Fig.~\ref{size}, we plot the maximum power of normalized ISI $\eta$ for different size of the RIS $\sqrt{N}$. We can observe that $\eta$ will decrease with a larger size of the RIS since it can provide more diversity to optimize. Moreover, $\eta$ will be lower with a higher reflection coefficient $\Gamma$. Under the assumption that phase shifts at the RIS are continuous, a larger reflection coefficient will provide more options on the amplitude of reflection rays, and thus can achieve a better performance. 

\vspace{-2mm}
\section{Conclusion}
In this paper, we have proposed the introduction of controllable paths artificially to mitigate multi-path fading via an RIS. As such, equalization can be done before signal reception. To eliminate ISI for multiple users, we have formulated a phase shift optimization problem and proposed an iterative algorithm to solve it. From simulation analysis, we can draw the following conclusions: 1) The proposed RIS-based spatial filter can effectively reduce the ISI. Even with 2-bit quantization, the performance of the proposed scheme is still better than that without the RIS; 2) The ISI will be further reduced with a larger RIS.



\vspace{-2mm}
\section{Appendix}
	According to the definition in (\ref{definition}), we have
	\begin{equation}
		\begin{aligned}
				\vspace{-1mm}
		&\nabla_{\theta_n}\hspace{-1mm} L(\theta_n^l,\eta^l,\mu_k^l) \hspace{-1mm}= \hspace{-1mm}\sum\limits_{k \in \mathcal{K}}\hspace{-1mm}\mu_k\nabla_{\theta_n}\hspace{-1mm}\left(\frac{Y_k(0)}{T} \hspace{-1mm}-\hspace{-1mm} y_k(0)\right)\hspace{-1mm}\left(\frac{Y^{*}_k(0)}{T} \hspace{-1mm}-\hspace{-1mm} y^{*}_k(0)\right)\\
		& = \hspace{-1mm}\sum\limits_{k \in \mathcal{K}}\hspace{-1mm}\mu_k\hspace{-1mm}\left(\nabla_{\theta_n}\hspace{-1mm}\frac{Y_k(0)Y^{*}_k(0)}{T^2} \hspace{-1mm}-\hspace{-1mm} \nabla_{\theta_n}\hspace{-1mm}\frac{Y_k(0)}{T}y^{*}_k(0) \hspace{-1mm}-\hspace{-1mm} \nabla_{\theta_n}\hspace{-1mm}y_k(0)\frac{Y^{*}_k(0)}{T}\right)
			\vspace{-1mm}
		\end{aligned}
	\end{equation}
	With the definitions of $A_{k,n}$ and $B_{k,n}$, we have
	\begin{equation}
		\begin{aligned}
				\vspace{-1mm}
		&\nabla_{\theta_n}\hspace{-1mm}Y_k(0)Y^{*}_k(0) = A_{k,n}B_{k,n}^{*}je^{j\theta_n^l} \hspace{-1mm}-\hspace{-1mm} B_{k,n}A_{k,n}^{*}je^{-j\theta_n^l},\\
		&\nabla_{\theta_n}\hspace{-1mm}Y_k(0)y^{*}_k(0) = - y_k^{*}(0)B_{k,n}je^{-j\theta_n^l},\\
		&\nabla_{\theta_n}\hspace{-1mm}y_k(0)Y^{*}_k(0) = y_k(0)B^{*}_{k,n}j e^{j\theta_n^l}.
			\vspace{-1mm}
		\end{aligned}
	\end{equation} 
This ends the proof.
	
\bibliographystyle{IEEEbib}
\bibliography{refs}

\end{document}